\documentclass[prd,showpacs]{revtex4}

\usepackage{amsmath}
\usepackage{amsfonts}
\usepackage{amsthm}
\usepackage{amssymb}
\usepackage{amscd}
\usepackage[british]{babel}
\usepackage{graphicx}
\usepackage{psfrag}
\usepackage{epsfig}
\usepackage{rotating}
\usepackage{times}

\theoremstyle{plain}

\newcommand{\boxend}{\flushright{$\Box$}}

\newcommand{\N}{{\mathbb N}}               
\newcommand{\Z}{{\mathbb Z}}               
\newcommand{\R}{{\mathbb R}}               

\renewcommand{\Im}{\mbox{\rm Im}}

\newcommand{\h}{\hbar}
\renewcommand{\tilde}{\widetilde}

\begin{document}

\title{
Effective gravity formulation that avoids singularities in quantum FRW cosmologies}

\author{Jaume Haro$^{1,}$\footnote{E-mail: jaime.haro@upc.edu} and
Emilio Elizalde$^{2,}$\footnote{E-mail: elizalde@ieec.uab.es,
elizalde@math.mit.edu}}

\affiliation{$^1$Departament de Matem\`atica Aplicada I, Universitat
Polit\`ecnica de Catalunya, Diagonal 647, 08028 Barcelona, Spain \\
$^2$Instituto de Ciencias del Espacio (CSIC) and Institut
d'Estudis Espacials de Catalunya (IEEC/CSIC), Facultat
de Ci\`encies \\ Universitat Aut\`onoma de Barcelona,
 Torre C5-Parell-2a planta, 08193 Bellaterra
(Barcelona) Spain}

\begin{abstract}

Assuming that time exists, a new, effective formulation of gravity is
introduced, which lies in between the Wheeler-DeWitt approach and ordinary
QFT. Remarkably, the Penrose-Hawking singularity of usual Friedman-Robertson-Walker
cosmologies is naturally avoided there. The theory is made explicit via specific examples,
and compared with loop quantum cosmology. It is
argued that it is the regularization of the classical Hamiltonian
performed in this last theory what avoid the singularity, rather than
quantum effects as in our case.

\end{abstract}


\maketitle



\section{Introduction}

In accordance with the
celebrated Penrose-Hawking singularity theorem, Friedman-Robertson-Walker (FRW)
cosmologies give rise to a singularity when the
strong energy condition holds, $\rho+3p>0$, $\rho$ being energy
density and $p$ pressure \cite{m-pv99}. A simple way to avoid this
singularity is introducing a scalar field that breaks the strong
energy condition \cite{em04,emt03,mtle05}. Another possibility is to
consider quantum effects due to vacuum polarization, as the one due
to a massless scalar field conformally coupled with gravity
\cite{pf73,d77,s80}. Remarkably, there is a very natural and fundamental alternative: to depart from
the Wheeler-DeWitt equation, $\hat{H}\Phi=0$ \cite{dw67,t01}, with $\hat{H}$ the
quantum Hamiltonian, by assuming that time has an absolute
meaning. The question of a possible singularity is
then addressed in terms of a Schr\"odinger equation with additional
conditions, what specifically defines our theory:
\begin{eqnarray}
i\hbar\partial_t\Phi(t)=\hat{H}\Phi(t),\ \,  \Phi(t^*)=\Psi, \ \,
\langle\hat{H}\rangle_{\Psi}=0, \ ||\Psi||=1. \label{1a}
\end{eqnarray}
We thus plainly assume the choice of a time direction
to have a physical solution, what goes against common lore. 
In this sense our approach is absolutely original,
revolutionary and, as we will prove, clarifying and predictive.

Some basic technical details on what will be demonstrated below.
The quantum Hamiltonian  $\hat{H}$, obtained with the usual rules of
quantum mechanics, is generically symmetric but not self-adjoint.
By von Neumann's theorem \cite{vgt07,bfg01}, it can be extended
to a self-adjoint operator
(sometimes in infinitely many ways).
Stone's theorem then applies, leading to a solution valid at all
time $t$ and, consequently, we can compute the average of the
quantum operator $\hat{a}$ corresponding to the classical scale
factor $a$. That is, we compute the following effective scale factor
$a_{eff}(t)\equiv\langle\Phi(t)|\hat{a}\Phi(t)\rangle$, where
$\Phi(t)$ is the solution of the effective Schr\"odinger equation
above. It is not difficult to see that if $\Phi(t)$ belongs to the
domain of the operator $\hat{a}$ at any time, then the effective
scale factor
is always strictly positive, and we can conclude that the
singularity is avoided. Physically, the self-adjoint extension of
the Hamiltonian operators that appears in FRW cosmologies can be
understood assuming that there is an infinitely barrier potential at
the point $a=0$, then when the effective factor scales approaches to
zero, at some finite time, it bounces and grows.

However to compute averages one usually works in Heisenberg's
picture, so $\dot{\hat{A}}=\frac{i}{\hbar}[\hat{H},\hat{A}]$, for
$\hat{A}$ any operator involved in the calculation. But, using this
formula one turns out to obtain, at some finite time, a negative
value for the average of the scalar factor operator. Such a
contradictory result can be explained by the fact that, at some
finite time, the commutator between the Hamiltonian and the operator
$\hat{A}$ is not well defined, what invalidates the final result
(physically one can explain this taking into account that in
Heisenberg picture the boundary conditions do not appear, i.e., the
barrier potential are not introduced and then the effective factor
scale has the freedom to take all the values in $\R$). That is, the
average of the scale factor is positive, but we do not have any
method to obtain an analytic information about its behavior, because
the Heisenberg picture fails to work, and it also turns out to be
impossible to obtain an explicit solution of the effective
Schr\"odinger equation. To this end loop quantum cosmology (LQC)
will be invoked \cite{a07,b06}. In what follows, we present a simple
demonstration of the above approach and will explicitly see how this
theory avoids the singularity. It will be also shown that it is the
regularization of the classical Hamiltonian \cite{rs93,t98,abl03}
what avoid the singularity, rather than quantum effects.

In the first of the three Appendix in the paper
 we present a brief mathematical review about the theory
of self-adjoint extensions of symmetric operators based on Von
Neumann's theorem. In the second one, we apply the effective
formulation to the case of a barotropic fluid where on can see
clearly the physical meaning of the self-adjoint extension of a
symmetric operator. As  specific examples, the dust and radiation
cases are treated in detail, showing that the self-adjoint extensions
of the respective Hamiltonian operators can be understood assuming
that there is an infinitely barrier of potential at $a=0$. Finally
in the last one, we show (resp. review) how to derive the standard
quantum fields theory in curved space-time from the effective
formulation (resp. the Wheeler-DeWitt equation). We also obtain,
from the effective formulation, the semi-classical Einstein
equation, that is, the back-reaction equation.

\section{The problem}
In this Section we consider an homogeneous and isotropic
gravitational field minimally coupled to an homogeneous scalar
field, which Lagrangian is given by \cite{mtle05}
\begin{eqnarray}\label{a1}
L(t)=\frac{3c^2}{8\pi G}(c^2
k-\dot{a}^2)a+\frac{1}{2}\dot{\phi}^2a^3-V(\phi)a^3,
\end{eqnarray}
where $G$ is Newton's constant and $k$ the three-dimensional
curvature. We are interested in the case $k=0$ and $V\equiv 0$,
previously studied in \cite{aps06,aps06a} within the framework of
LQC. This interest comes from the fact that in the chaotic
inflationary model with $V=\frac{1}{2}m^2\phi^2$, at very early
times before the inflationary period, one has $\dot{\phi}\gg V$ (see
for details \cite{mu05}). Then the potential can be neglected and
one has $\rho\cong p$. Consequently this model give rise to a
singularity at very early times, that we want to avoid using the
effective formulation described in the Introduction.

Defining the angle variable $\psi$ by $
 \phi=\sqrt{3c^2/4\pi G}\,{\psi},
$ the Lagrangian becomes ($l_p$ denotes the Planck length)
\begin{eqnarray}\label{a3}
L=-\frac{\gamma^2}{2}(\dot{a}^2a-\dot{{\psi}}^2a^3),\quad
\mbox{with}\quad \gamma^2=\frac{3c^2}{4\pi G}=\frac{3\hbar}{4\pi
cl_p^2}.
\end{eqnarray}
Using the conjugate momenta, $p_a\equiv -\gamma^2\dot{a}a$ and
$p_{\psi}\equiv \gamma^2\dot{\psi}a^3$, we can write the Hamiltonian
as
\begin{eqnarray}\label{a4}
H=\frac{1}{2\gamma^2a^3}\left[-(ap_a)^2+p_{\psi}^2\right].
\end{eqnarray}
The classical dynamic equations are
\begin{eqnarray}\label{a5}
\dot{a}=-\frac{1}{\gamma^2a^2}ap_a,\, \dot{(ap_a)}=3H,\,
\dot{\psi}=\frac{1}{\gamma^2a^3}p_{\psi},\, \dot{p}_{\psi}=0,
\end{eqnarray}
together with the constraint $H=0$, that is $(ap_a)^2=p_{\psi}^2$.
Integrating $(\ref{a5})$  we obtain the following solution
\begin{eqnarray}\label{a6}
&& 0>a(t)p_{a}(t)\equiv p_{\alpha}^*,\quad
p_{\psi;\pm}(t)\equiv p_{\psi,\pm}^*=\pm p_{\alpha}^*,\\
&& \hspace*{-3mm} {a}(t)=a^*\left[1- \frac{3\,
p_{\alpha}^*(t-t^*)}{\gamma^2(a^*)^3}\right]^{1/3} \hspace*{-3mm}, \
\ {{\psi}_{\pm}}(t)=\mp\ln\frac{a(t)}{a^*} +{\psi}^*. \nonumber
\end{eqnarray}
Note that the solution $({a}(t),{{\psi}_{\pm}}(t))$ is defined in
the interval $(t_{s},+\infty)$, where $
t_{s}=t^*+(\gamma^2/3p_{\alpha}^*)\, (a^*)^3. $ At this time we have
$ {a}(t_{s})=0$, and ${\psi}_{\pm}(t_{s})=\pm\infty, $ that is, the
dynamics is singular at $t=t_s$. Note that we can write
${a}(t)={a}^*\left[1-(t-t^*)/(t_{s}-t^*)\right]^{1/3}.$ Finally, we
see that from Eqs.~$(\ref{a5})$   we have $d{\ln
a}/d{\psi}_{\pm}=\mp 1$, and
  conclude:
\begin{eqnarray}\label{a9}
{a}={a}^*e^{\mp(\psi_{\pm}-\psi^*)}.
\end{eqnarray}

\subsection{Quantum dynamics}
We now use the  quantization rule:
\begin{eqnarray}\label{a10}
g^{AB}p_Ap_B\longrightarrow
-\hbar^2\nabla_A\nabla^A=-\frac{\hbar^2}{\sqrt{|g|}}\partial_A(\sqrt{|g|}g^{AB}\partial_B),
\end{eqnarray}
and obtain the quantum Hamiltonian
\begin{eqnarray}\label{a11}
\hat{H}\equiv\frac{\hbar^2}{2\gamma^2a^3}(a\partial_aa\partial_a-\partial^2_{\psi^2}).
\end{eqnarray}
Introducing the operators $ \widehat{ap_a}\equiv-i\hbar
a^{-1/2}\partial_aa^{3/2};\quad \hat{p}_{\psi}\equiv
-i\hbar\partial_{\psi}, $ we can write $
\hat{H}\equiv\frac{1}{2\gamma^2}a^{-3/2}[\hat{p}_{\psi}^2-(\widehat{ap_a})^2]a^{-3/2}.
$ The dynamical equations  in the Heisenberg picture are
\begin{eqnarray}\label{a14}
&&\frac{d\hat{a}}{dt}
=-\frac{1}{\gamma^2\hat{a}^2}(\widehat{ap}_{a}+i\hbar),\quad
\frac{d(\widehat{ap}_{a})}{dt} =3\hat{H},\nonumber \\ &&
\frac{d\hat{\psi}}{dt}=\frac{1}{\gamma^2\hat{a}^3}\hat{p}_{\psi},
\quad\frac{d\hat{p}_{\psi}}{dt}=0,
\end{eqnarray}
with $ \langle \Phi|\Psi\rangle\equiv  \int_0^{\infty}da\int_{\R}
d\psi \Phi^*(a,\psi)a^{2}\Psi(a,\psi) $ as inner product and
operator average  $ \langle \hat{A}\rangle_{\Psi}\equiv \langle
\Psi|\hat{A}\Psi\rangle, $ $||\Psi||=1$. \medskip

\noindent {\sl Example 1.} Consider the wave-function
($|p_{\alpha}^*|=|p_{\psi}^*|$): $ \Psi(a,\psi)\equiv
\frac{a^{-3/2}}{(\sigma\pi)^{1/2}}e^{-\frac{\ln^2(a/\bar{a})}{2\sigma}}
e^{-\frac{(\psi-\psi^*)^2}{2\sigma}}e^{\frac{i}{\hbar}(\ln(a/\bar{a})
p_{\alpha}^*+\psi p_{\psi}^*)}. $ Then:
$\langle\hat{a}\rangle_{\Psi}=\bar{a}e^{\sigma/4}\equiv a^*$,
$\langle\hat{\psi}\rangle_{\Psi}={\psi}^*$,
$\langle\widehat{ap_a}\rangle_{\Psi}={p}_{\alpha}^*$,
$\langle\hat{p}_{\psi}\rangle_{\Psi}={p}_{\psi}^*$, and
$\langle\hat{H}\rangle_{\Psi}=0$.

\subsection{The Wheeler-DeWitt equation}

Compare at this point with the Wheeler-DeWitt equation paradigm
(WDW) \cite{dw67}
\begin{eqnarray}\label{a18}
\hat{H}\Phi=0\longrightarrow
(a\partial_{a}a\partial_{a}-\partial^2_{\psi^2})\Phi=0,
\end{eqnarray}
with general solution ($\tilde{a}$ is a length constant)
\begin{eqnarray}\label{a19}
\Phi(a,\psi)=f_+(\ln(a/\tilde{a})+\psi)+f_-(\ln(a/\tilde{a})-\psi).
\end{eqnarray}
The quantum version of Eq.~$(\ref{a9})$ is $
\Phi_+(a,\psi)=f_+(\ln(a/{a}^*)+\psi-\psi^*),$ $
\Phi_-(a,\psi)=f_-(\ln(a/{a}^*)-\psi+\psi^*), $ with $f_{\pm}$ a
function picked around $0$, as for instance
$f_{\pm}(z)=e^{-z^2/\sigma}$. From this result we see that the wave
is always picked around the classical solution, but we cannot
conclude that its dynamical behavior is singular since here time
does {\it not} appear.

In order to understand the dynamics, we postulate the effective
equation (\ref{1a}). Note that $\langle\hat{H}\rangle_{\Psi}=0$
implies $\langle\hat{H}\rangle_{\Phi(t)}=0 $, and $||\Psi||=1$
implies $||\Phi(t)||=1$, $\forall t\in \R$. Then, if the solution of
the problem exists for all $t$, it is easy to prove that the
effective scalar factor,
$a_{eff}(t)\equiv\langle\hat{a}\rangle_{\Phi(t)}$, never vanishes.
In fact, the condition $||\Psi||=1$ implies $||\Phi(t)||=1$, i.e.,
$\int_0^{\infty}da\int_{\R} d\psi a^2|\Phi(t,a,\psi)|^2=1$, and thus
we have: $a_{eff}(t)=\langle\hat{a}\rangle_{\Phi(t)}=
\int_0^{\infty}da\int_{\R} d\psi a^3|\Phi(t,a,\psi)|^2\not=0$.

To do the calculation, we consider the quantity
$\langle\hat{a}^3\rangle_{\Phi(t)}$. We have
$\frac{d}{dt}\langle\hat{a}^3\rangle_{\Phi(t)}=\frac{i}{\hbar}\langle[\hat{H},\hat{a}^3]\rangle_{\Phi(t)}=
-\frac{3}{\gamma^2}\langle\widehat{ap_a}\rangle_{\Phi(t)}$ and
$\frac{d^2}{dt^2}\langle\hat{a}^3\rangle_{\Phi(t)}=-\frac{3i}{\hbar\gamma^2}\langle[\hat{H},\widehat{ap_a}]\rangle_{\Phi(t)}=
-\frac{9}{\gamma^2}\langle\hat{H}\rangle_{\Phi(t)}=0,$ due to the
remark above. Consequently, we obtain
\begin{eqnarray}\label{a22}
\langle\hat{a}^3\rangle_{\Phi(t)}=\langle\hat{a}^3\rangle_{\Psi}-\frac{3}{\gamma^2}\langle\widehat{ap_a}\rangle_{\Psi}(t-t^*).
\end{eqnarray}
For the function of Example 1, we get
\begin{eqnarray}\label{a23}
\langle\hat{a}^3\rangle_{\Phi(t)}=\bar{a}^3e^{9\sigma/4}-\frac{3}{\gamma^2}p_{\alpha}^*(t-t^*),
\end{eqnarray}
and this contradicts the fact that
$\langle\hat{a}^3\rangle_{\Phi(t)}\geq 0, \forall t\in \R$. However,
since the operator $\hat{H}$ is symmetric and real, using von
Neumann's theorem \cite{vgt07,bfg01} it can be extended to a
self-adjoint one, and then the solution of the problem (\ref{1a})
exists here for any $t$ (Stone's theorem). From this result, we
conclude that there is a value of $t$ for which some of the
commutators $[\hat{H},\hat{a}^3]$ and/or $[\hat{H},\widehat{ap_a}]$
do not exist; thus the final result (\ref{a23}) is incorrect. The
drawback of this method is the lack of an analytic procedure to
calculate the average since, in general, there is no explicit
formula that gives information on the regular behavior of the
average of the scale factor operator. Fortunately, a useful way
exists to directly analyze the singularity, namely loop quantum
cosmology (LQC). Before using it in our problem, we consider another
example where the above contradictions can be easily depicted.
\medskip

\noindent {\sl Example 2.} Consider now the problem
\begin{eqnarray}\label{b}
i\partial_t\Phi=-i\hbar c\partial_x\Phi\equiv c\hat{p}\Phi, \
\forall x\in [0,2\pi],\quad \Phi(0)=\Psi,
\end{eqnarray}
being $\hat{p}$ self-adjoint in the domain \cite{bfg01}: $
D_{\hat{p}}=\{\Psi $ absolutely continuous in $ [0,2\pi],
\partial_x\Psi\in \mathcal{L}^2[0,2\pi], \\ \Psi(0)=\Psi(2\pi)\}$. Let
$\Phi(t)$ be the solution of our effective formulation (\ref{b}). We
want to calculate $\langle\hat{x}\rangle_{\Phi(t)}\equiv
\int_0^{2\pi}x|\Phi(t,x)|^2dx$. Using $[\hat{p},\hat{x}]=-i\hbar$,
we get
 $\langle\dot{\hat{x}}\rangle_{\Phi(t)}=c$, i.e.,
$\langle\hat{x}\rangle_{\Phi(t)}=\langle\hat{x}\rangle_{\Psi}+ct$
which is {\it not} positive $\forall t$. What actually happens is
that, for some  $t$, we have $\hat{x}\Phi(t)\notin D_{\hat{p}}$, and
then $\hat{p}\hat{x}\Phi(t)$ has no sense,  neither the formula
  $\langle\dot{\hat{x}}\rangle_{\Phi(t)}=\frac{i}{\hbar}\langle[c\hat{p},{\hat{x}}]\rangle_{\Phi(t)}$.
To see this in detail, consider the initial state
\begin{eqnarray}\Psi(x)=\sqrt{\frac{3}{2\pi^2}}\left\{\begin{array}{ccc}
x,&\mbox{for}&x\in[0,\pi],\\
2\pi-x,&\mbox{for}&x\in[\pi,2\pi].
\end{array}\right.\end{eqnarray}
Fourier analysis provides the following solution of the
Schr\"odinger equation $$
\Phi(t,x)=\sqrt{\frac{3}{2\pi^2}}\left\{\frac{{\pi}}{2}-\frac{4}{\pi}
\sum_{n\in\Z}
\frac{1}{(2n+1)^2}\cos\left[(2n+1)(x-ct)\right]\right\}. $$ Then, at
$t=0$ we have $x\Phi(0,x)\in D_{\hat{p}}$ but if we choose $t=\pi/c$
we obtain $x\Phi(\pi/c,x)_{|_{x=0}}=0$, and
$x\Phi(\pi/c,x)_{|_{x=2\pi}}=\sqrt{6}\pi/c$, what means effectively
that $x\Phi(\pi/c,x)\not\in D_{\hat{p}}$. However, note that
$\langle\hat{x}\rangle_{\Phi(t)}$ exists for all $t\in\R$, its value
being: $$0<
\langle\hat{x}\rangle_{\Phi(t)}=\int_0^{2\pi}\frac{3x}{2\pi^2}\left\{
\frac{\pi}{2}-\frac{4}{\pi}\sum_{n\in\Z}\frac{1}{(2n+1)^2}
\cos\left[(2n+1)(x-ct)\right]\right\}^2dx<2\pi. $$

\section{Loop Quantum Cosmology to rescue}

We shall now involve (a simplified version of)
 LQC (for a rigorous formulation see \cite{b06,abl03}), with different
 variables and a different quantum space of states, adapted to
make contact with our theory above.

Consider the variables $p\equiv a^2$ and $x\equiv \dot{a}$. Their
Poisson bracket is $ \{x,p\}=\frac{8\pi G}{3c^2}=\frac{2}{\gamma^2}.
$ We also consider the holonomies $ h_j(n)\equiv e^{-i\frac{n\iota
x}{2c}\sigma_j} =\cos(\frac{n\iota
x}{2c})-i\sigma_j\sin(\frac{n\iota x}{2c})$ \cite{s06}, where
$\sigma_j$ are the Pauli matrices and $\iota$ is the Barbero-Immirzi
parameter. We easily obtain Ashtekar-Barbero's formula \cite{s06}
\begin{eqnarray}\label{a25}
a^{-1}=\frac{-i\hbar}{4\pi l_p^2\iota}Tr
\sum_{j=1}^3\sigma_jh_j(1)\{h^{-1}_j(1),a\}.
\end{eqnarray}
To get the gravitational part of the Hamiltonian, we cannot directly
use this one: $H_{grav}=-\frac{3c^2}{8\pi G}x^2 \sqrt{p}$, which
leads to singular classical dynamics. We may use the general
formulae of loop quantum gravity (LQG) to obtain the regularized
Hamiltonian \cite{abl03,t01,aps06a}: $$
H_{grav,\iota}\equiv-\frac{\hbar^2c}{32\pi^2l_p^4\iota^3}
\sum_{i,j,k}\varepsilon^{ijk}Tr\left[
h_i(1)h_j(1)h_i^{-1}(1)h_j^{-1}(1)h_k(1)\{h_k^{-1}(1),a^3\}\right]
=-\frac{\gamma^2c^2}{2\iota^2}a\sin^2\frac{\iota x}{c}, $$ which is
bounded when the extrinsic curvature $x/2$ (a half of the velocity
of the scalar factor) diverges, and approaches $ H_{grav}$ for small
values of
$x$. Then, taking this regularized Hamiltonian as the gravitational
part of the full one, this last is given by \cite{sv05,s06}
\begin{eqnarray}\label{a27}
{{H}}_{\iota}\equiv -\frac{\gamma^2c^2}{2\iota^2}a\sin^2\frac{\iota
x}{c} +\frac{1}{2\gamma^2} a^{-3}{p}_{\psi}^2,
\end{eqnarray}
and the dynamical equations are
\begin{eqnarray}\label{a28}
\dot{a}=\{a,{H}_{\iota}\}=\frac{c}{2\iota}\sin\frac{2\iota x}{c},\
\dot{x}=\{x,{H}_{\iota}\}=-\frac{2}{\gamma^2a^4}p_{\psi}^2.
\end{eqnarray}
Imposing the Hamiltonian constraint ${H}_{\iota}=0$, we obtain
\begin{eqnarray}\label{a29}
\dot{a}^2=\frac{p_{\psi}^2}{\gamma^4a^4}\left(1-
\frac{p_{\psi}^2\iota^2}{\gamma^4a^4c^2}\right),
\end{eqnarray}
and since $p_{\psi}\equiv p_{\psi}^*$ is constant, we get the
following bounce, $\dot{a}=0$ when
$a=\frac{1}{\gamma}\sqrt{\frac{p_{\psi}^*\iota}{c}}=2l_p\sqrt{\frac{\pi
p_{\psi}^*\iota}{3\hbar}}$. Consequently, there is no singularity
because the range of $a(t)$ is $\left[ 2l_p\sqrt{\frac{\pi
p_{\psi}^*\iota}{3\hbar}},+\infty\right), \forall t\in \R$. In fact,
at earlier times the scalar factor is very big, then it decreases,
and when it arrives at the turning value it increases forever.
 Moreover,
this solution yields a period of inflation \cite{b02}, namely, from
the Friedmann Eq.~(\ref{a29}): $\ddot{a}>0$, for $a\in
\left(2l_p\sqrt{\frac{\pi
p_{\psi}^*\iota}{3\hbar}},2l_p\sqrt{\frac{\sqrt{2}\pi
p_{\psi}^*\iota}{3\hbar}}\right)$. Finally, note that when $a\gg
l_p$, Eq.~(\ref{a29}) coincides with (\ref{a5}).

A different way to understand these features is to write
Eq.~$(\ref{a29})$ as $
\left(\frac{\dot{a}}{a}\right)^2=\frac{2}{\gamma^2}\rho_{eff}, $
where we have introduced the effective energy density $
\rho_{eff}\equiv\frac{p_{\psi}^2}{2a^6}\left(1-
\frac{p_{\psi}^2\iota^2}{\gamma^4a^4c^2}\right). $ Taking the
derivative, $
\dot{\rho}_{eff}=-3\left(\frac{\dot{a}}{a}\right)(\rho_{eff}+p_{eff}),
$ where the effective pressure is $
p_{eff}\equiv\frac{p_{\psi}^2}{2a^6}\left(1-
\frac{7p_{\psi}^2\iota^2}{3\gamma^4a^4c^2}\right). $ But then is
easy to see that the strong energy condition
${\rho}_{eff}+3{p}_{eff}>0$ is broken, when the scale factor lies in
the interval $\left(2l_p\sqrt{\frac{\pi
p_{\psi}^*\iota}{3\hbar}},2l_p\sqrt{\frac{\sqrt{2}\pi
p_{\psi}^*\iota}{3\hbar}}\right)$, consequently, the singularity is
avoided. Moreover, for $a\in\left(2l_p\sqrt{\frac{\pi
p_{\psi}^*\iota}{3\hbar}},2l_p\sqrt{\frac{\sqrt{5}\pi
p_{\psi}^*\iota}{3\sqrt{3}\hbar}}\right)$, there is a period of
super-inflation; that is, in this interval, one has
$\frac{p_{eff}}{\rho_{eff}}<-1$.

The following remark is in order.
Similar results are obtained in the case $k=1$ and $V\equiv 0$. Now
the classical Hamiltonian is given by
\begin{eqnarray}\label{b1}
H=\frac{1}{2\gamma^2a^3}(-(ap_a)^2+p_{\psi}^2-\gamma^4c^2 a^4),
\end{eqnarray}
and the regularized one, is
\begin{eqnarray}\label{b2}
{{H}}_{\iota}\equiv
-\frac{\gamma^2c^2}{2\iota^2}a\left(\sin^2\left(\frac{\iota
x}{c}\right)+\iota^2\right) +\frac{1}{2\gamma^2} a^{-3}{p}_{\psi}^2.
\end{eqnarray}
Then, using the Hamiltonian constraint ${{H}}_{\iota}=0$, is easy to
obtain the Friedmann equation
\begin{eqnarray}\label{b3}
\dot{a}^2=\left(\frac{p_{\psi}^2}{\gamma^4a^4}-c^2\right)\left(1+\iota^2-
\frac{p_{\psi}^2\iota^2}{\gamma^4a^4c^2}\right),
\end{eqnarray}
and since, $p_{\psi}\equiv p_{\psi}^*$ is constant and
$\dot{a}^2\geq 0$, we can deduce that $a\in\left[2l_p\sqrt{\frac{\pi
p_{\psi}^*\iota}{3\hbar\sqrt{1+\iota^2}}},2l_p\sqrt{\frac{\pi
p_{\psi}^*}{{3}\hbar}}\right]$, and clearly the singularity is
avoided. In this case, we have an oscillating universe.

 Another equivalent way to do this, is to use the variable
$\tilde{x}\equiv \dot{a}+c$, then following \cite{apsv07,skl07} we
obtain the regularized hamiltonian
\begin{eqnarray}\label{b4}
\tilde{{H}}_{\iota}\equiv
-\frac{\gamma^2c^2}{2\iota^2}a\left(\sin^2\left(\frac{\iota
(\tilde{x}-c)}{c}\right)-\sin^2(\iota)+2\iota^2\right)
+\frac{1}{2\gamma^2} a^{-3}{p}_{\psi}^2.
\end{eqnarray}
It is clear, that from this last regularized hamiltonian the scale
factor has the same behavior that described in equation (\ref{b3}).

\subsection{Quantization}

To quantize we perform the usual
 change $\{A,B\}\rightarrow
-\frac{i}{\hbar}[\hat{A},\hat{B}]$. Note that the system is
$\frac{4\pi c}{\iota}$-periodic with respect to the variable $x$,
thus we consider the space of $\frac{4\pi c}{\iota}$-periodic
functions  and introduce  the inner-product $
\langle\Psi|\Phi\rangle\equiv\int_{\R}d\psi\int_{-\frac{2\pi
c}{\iota}}^{\frac{2\pi c}{\iota}}dx \Psi^*(x,\psi)\Phi(x,\psi). $
Completion of this space with respect to this product is the space
of square-integrable functions in  $[-\frac{2\pi c}{\iota}
,\frac{2\pi c}{\iota} ]$. Note that, rigorously, the definition of
the Hilbert space is more complicated: ${\mathcal
L}^2(\R_{Bohr},d\mu_{Bohr})$ where $\R_{Bohr}$ is the
compactification of $\R$ and $\mu_{Bohr}$  the Haar measure on it
\cite{aps06a}. However, for our purposes the Hilbert space
${\mathcal L}^2[-\frac{2\pi c}{\iota} ,\frac{2\pi c}{\iota} ]$ will
suffice.

We quantize the variable $p$ as above and, using the fact that
$p>0$, we can define $
\hat{p}\equiv\left(-\frac{4\hbar^2}{\gamma^4}\partial^2_{x^2}\right)^{1/2},
$
 the volume operator
$\hat{V}\equiv \hat{p}^{3/2}$, and the scale factor
$\hat{a}\equiv\hat{p}^{1/2}$. The eigenfunctions of these operators
are $|n\rangle\equiv \sqrt{\frac{\iota}{4\pi
c}}{e^{\frac{in\iota}{2c}x}}$, and their eigenvalues are
$(\hat{p})_n=\frac{4\pi}{3}\iota |n|l_p^2$,
$(\hat{V})_n=\left(\frac{4\pi}{3}\iota |n|l_p^2\right)^{3/2}$ and
$(\hat{a})_n=\sqrt{\frac{4\pi}{3}\iota |n|}l_p$. Using
Eq.~(\ref{a25}),
\begin{eqnarray}\label{a40}
\hat{a}^{-1}\equiv -\frac{1}{4\pi l_p^2\iota}
Tr\sum_{j=1}^3\sigma_j\hat{h}_j(1)[\hat{h}^{-1}_j(1),\hat{V}^{1/3}],
\end{eqnarray}
and consequently the corresponding quantum operator is
\begin{eqnarray}\label{a41}
\hat{a}^{-1}|n\rangle= \sqrt{\frac{3}{4\pi \iota}} \frac{1}{l_p}
\left(\sqrt{|n+1|}-\sqrt{|n-1|}\right)|n\rangle,
\end{eqnarray}
whose eigenvalues, when $n\gg 1$, satisfy
$(\hat{a}^{-1})_n=1/(\hat{a})_n$.

The quantization of the gravitational part of the Hamiltonian,
depends on the order  we fix. For instance,  $
\hat{H}_{grav,\iota}\equiv \frac{i\hbar
c}{32\pi^2l_p^4\iota^3}\sum_{i,j,k}\varepsilon^{ijk}Tr\left[
\hat{h}_i^{-1}(1)\hat{h}_j^{-1}(1)\hat{h}_k(1)[\hat{h}_k^{-1}(1),\hat{V}]\hat{h}_i(1)\right.$
$\left. \hat{h}_j(1)\right]$ gives us a self-adjoint operator, or
the direct quantization of the expression
$-\frac{\gamma^2c^2}{2\iota^2}a\sin^2\frac{\iota x}{c}$ yields
\begin{eqnarray}\label{a44}
\widehat{\tilde{H}}_{grav,\iota}\equiv -\frac{\gamma^2c^2}{2
\iota^2}\hat{a}^{1/2}\sin^2\left(\frac{\iota
x}{c}\right)\hat{a}^{1/2}.
\end{eqnarray}
If we use this operator (\ref{a44}) as the gravitational part of the
full Hamiltonian, then this is given by
\begin{eqnarray}\label{a45}
\widehat{\tilde{H}}_{\iota}\equiv -\frac{\gamma^2 c^2}{2
\iota^2}\hat{a}^{1/2}\sin^2\left(\frac{\iota
x}{c}\right)\hat{a}^{1/2} +\frac{1}{2\gamma^2}
\left(\hat{a}^{-1}\right)^3\hat{p}_{\psi}^2,
\end{eqnarray}
and in this case the WDW equation becomes
$\widehat{\tilde{H}}_{\iota}\Phi=0$ which,  expanding $\Phi$ as
$\Phi=\sum_{n\in\N}\Phi_n(\psi)|n\rangle$, turns into
\begin{eqnarray}
&& \hspace*{-8mm}
2\sqrt{|n|}\Phi_n-|n(n-4)|^{1/4}\Phi_{n-4}-|n(n+4)|^{1/4}\Phi_{n+4}\nonumber
\\ && \hspace*{-4mm}
+4\left(\sqrt{|n+1|}-\sqrt{|n-1|}\right)^3\partial^2_{\psi^2}\Phi_n=0,
\ \ n\in\N. \label{a46}
\end{eqnarray}

Summing up,  the effective equation
$i\hbar\partial_t\Phi=\widehat{\tilde{H}}_{\iota}\Phi$ with the
condition $\langle\widehat{\tilde{H}}_{\iota}\rangle_{\Phi(t)}=0$
yields an average of the scalar factor operator that has essentially
the same behavior as the classical solution of Eq.~(\ref{a29}). This
owes to the fact that the domain of the holonomy operators is the
whole space, so that one can safely use the Heisenberg picture in
order to obtain the quantum version of the classical equations. This
gives generically small corrections to the classical behavior.

A final remark is in order. The singularity is avoided in the
classical theory after regularization of the Hamiltonian.
Quantization of this new Hamiltonian provides then a self-adjoint
operator. It is important to realize that it is the regularization
of the classical Hamiltonian what avoids the singularity, rather
than the quantum effects. This is overlooked in some papers, where
it is claimed that  quantum effects are essential to
 avoid the big bang singularity
\cite{aps06a,b01,b01a}. Note that in these approximations one
already starts from the quantum theory and then, using the quantum
operators an effective Hamiltonian is obtained \cite{v05,dh04,bd05}
which, in fact, is in essence the Hamiltonian (\ref{a27}). This is
maybe the reason why it is plainly concluded there that quantum
effects, provided by LQC, are responsible for avoiding the big bang
singularity. Here, with our alternative formulation we have shown,
by means of explicit examples, that this need not be the case.

\section{Conclusions}
We have presented here an effective formulation that
naturally avoids the big bang singularity: in essence
Schr\"odinger's equation with the condition that the average of the
Hamiltonian operator be zero. This is different from the
Wheeler-DeWitt equation where one impose that the Hamiltonian
operator annihilates the wave-function, and the arrow of time is yet to be selected.
In our theory, physical time has essentially the same
meaning as in the classical theory, and the relevant quantities
are averages of quantum operators, as e.g. the average
of the scale factor operator---which is by definition strictly
positive---and {\it no singularity appears at finite time}.
Our approach is remarkably natural (once time is assumed to exist),
revolutionary and predictive, albeit rather non-trivial. It does not seem easy to
produce an analytic formula that provides information on the
behavior of the observable averages. Only numerical results look
feasible at this point.

Another way to deal with the classical big bang singularity is LQC.
We have here involved a simplified version of this theory and shown
that, in contradistinction with the theory presented above, in LQC
it is the regularization of the classical Hamiltonian that seems to
avoid the singularity, and not the quantum effects obtained after
quantization of the regularized Hamiltonian.

\section{Appendix A: Self-adjoint extensions of symmetric operators}

In this mathematical Appendix we present a brief review of the
theory of the self-adjoint extensions of symmetric operators.

Let $\hat{A}$ be a linear operator that is defined on a dense subset
$D_{\hat{A}}$ of a separable Hilbert space $\mathcal H$. The adjoint
$\hat{A}^{\dagger}$ of $\hat{A}$ is defined on those vectors
$\Phi\in \mathcal H$ for which there exist  $\tilde{\Phi}\in
\mathcal H$ such that $\langle\Phi|\hat{A}\Psi\rangle=
\langle\tilde{\Phi}|\Psi\rangle\quad \forall\Psi\in D_{\hat{A}}$,
and $\hat{A}^{\dagger}$ is defined on such $\Phi$ as
$\hat{A}^{\dagger}\Phi\equiv\tilde{\Phi}$.

The graph of an operator $\hat{A}$ is a subset of $\mathcal
H\bigoplus\mathcal H$, defined by $G_{\hat{A}}\equiv
\{(\Phi,\hat{A}\Phi); \Phi\in D_{\hat{A}}\}$, and $\hat{A}$ is
called closed, which is written as $\bar{\hat{A}}=\hat{A}$, if its
graph is a closed set. 
An extension of an operator $\hat{A}$, namely $\hat{A}_{ext}$, is an
operator that satisfies $D_{\hat{A}}\subset D_{\hat{A}_{ext}}$ and
$\hat{A}_{ext}\Phi=\hat{A}\Phi\quad \forall\Phi\in D_{\hat{A}}$.

An operator $\hat{A}$ is symmetric if
$\langle\Phi|\hat{A}\Psi\rangle=\langle\hat{A}\Phi|\Psi\rangle\quad
\forall \Psi,\Phi\in D_{\hat{A}}$. Then, a symmetric operator
$\hat{A}$ always admits a closure (a minimal closed extension),
which is its double adjoint, i.e.,
$\bar{\hat{A}}=\hat{A}^{\dagger\dagger}$.
The adjoint of a symmetric operator $\hat{A}$ is always a closed
extension of it,
and it is self-adjoint when $D_{\hat{A}}=D_{\hat{A}^{\dagger}}$.
The deficiency subspaces $\mathcal{N_{\pm}}$ of the operator
$\hat{A}$ are defined by
\begin{eqnarray}\label{a30}
\mathcal{N_{\pm}} =\left\{\Phi\in D_{\hat{A}^{\dagger}},\quad
\hat{A}^{\dagger}\Phi=z_{\pm}\Phi,\quad \pm\Im(z_{\pm})>0\right \},
\end{eqnarray} and the deficiency
indices $n_{\pm}$ of $\hat{A}$  are its dimensions. Note that, these
two definitions do not depend on the values of $z_{\pm}$.

The following theorem is due to Von Neumann:

For a closed symmetric operator $\hat{A}$ with deficiency indices
$n_{\pm}$ there are three possibilities: 

\noindent a) If $n_+=n_-=0$, then $\hat{A}$ is self-adjoint.

\noindent b) If $n_+=n_-=n\geq 1$, then $\hat{A}$ has infinitely many
self-adjoint extensions parametrized by an unitary $n\times n$
matrix. Each unitary matrix $U_n:\mathcal{N_{+}}\rightarrow
\mathcal{N_{-}}$, characterizes a self-adjoint extension
$\hat{A}_{U_n}$ as the restriction of $\hat{A}^{\dagger}$ to the
domain
\begin{eqnarray*}
D_{\hat{A}_{U_n}}=\{\Phi+\Phi_{z_+}+U_n\Phi_{z_+}; \Phi\in
D_{\hat{A}}\quad \Phi_{z_+}\in\mathcal{N_{+}}\}.
\end{eqnarray*}

\noindent c) If $n_+\not=n_-$, then $\hat{A}$ has no  self-adjoint extensions.

\section{Appendix B: Effective formulation for a barotropic perfect  fluid}

In this Appendix we apply our effective formulation to the case of a
barotropic perfect fluid with state equation $p=\omega\rho$. The
Lagrangian of the system  in the flat case ($k=0$) is
\begin{eqnarray}\label{a31}
L=-\frac{\gamma^2}{2}\dot{a}^2a-\rho(a)a^3.
\end{eqnarray}
The momentum and the Hamiltonian are respectively
$p_a=-\gamma^2\dot{a}a$, and $H=-\frac{1}{2\gamma^2
a}p_a^2+\rho(a)a^3$. Using the conservation equation
$\dot{a}=-3H(\rho+p)$ we have
$\rho(a)=\rho_0\left({a}/{a_0}\right)^{-3(\omega+1)}$, then the
dynamical equations become
\begin{eqnarray}\label{a32}
\dot{a}=-\frac{p_a}{\gamma^2 a};\quad
\dot{p_a}=-\frac{p_a^2}{2\gamma^2 a^2}+3\omega\rho(a) a^2,
\end{eqnarray}
with the constraint $H=0$.

The quantization rule (\ref{a10}) give us the following Hamiltonian
operator
\begin{eqnarray}\label{a33}
\hat{H}=\frac{\h^2}{2\gamma^2a}\partial^2_{a}+\rho(a)a^3,
\end{eqnarray}
which is symmetric with respect to the inner product
$\langle\Phi|\Psi\rangle=\int_0^{\infty}da a\Phi^*(a)\Psi(a)$ of the
Hilbert space ${\mathcal L}^2((0,\infty),ada)$.

To apply the theory presented in the Appendix A, first we consider
the case $\omega=0$ (dust matter), whose Hamiltonian is
$\hat{H}=\frac{\h^2}{2\gamma^2a}\partial^2_{a}+\rho_0a_0^3$. To
study the self-adjoint extensions of this operator we need to
determine the deficiency subspaces $\mathcal{N_{\pm}}$,
that is, we must solve the equation $\hat{H}\Phi=z_{\pm}\Phi$ with
$||\Phi||<\infty$.
Since the definition of these spaces do not depend on $z_{\pm}$, we
choose, $z_{\pm}=\pm i\rho_0a_0^3$. Then the solutions of
$\hat{H}\Phi=\pm i\rho_0a_0^3\Phi$ are the Airy's functions
$\Phi_{1,\pm}\equiv Ai(\beta_{\pm} a)$ and $\Phi_{2,\pm}\equiv
Bi(\beta_{\pm} a)$, where $\beta_{\pm}\equiv
\left(\frac{4\gamma^2\rho_0a_0^3}{\hbar^2}\right)^{1/3}e^{\pm
i\pi/4}$ \cite{as72}. However, only $\Phi_{1,\pm}$ has finite norm,
and then both spaces has dimension $1$. Von Neumann's theorem  says
us that $\hat{H}$ has infinitely many self-adjoint extensions,
namely $\hat{H}_{SA}$, parametrized by an unitary $1\times 1$
matrix, i.e., by $e^{\alpha}$ being $\alpha\in\R$. To obtain an
explicit expression of the domain of these self-adjoint extensions
we must impose \cite{vgt07,acp03} $\langle
\hat{H}_{SA}(\Phi_++e^{i\alpha}\Phi_-)|\Psi\rangle= \langle
\Phi_++e^{i\alpha}\Phi_-|\hat{H}_{SA}\Psi\rangle \quad \forall
\Psi\in D_{\hat{H}_{SA}}$. It is not difficult to show that, this
condition is accomplished when
\begin{eqnarray}\label{a34}
\frac{\Psi(0)}{\Psi'(0)}=\frac{Ai(0)}{Ai'(0)|\beta_{+}|}\frac{1}{1+\tan(\alpha/2)}\equiv
r,\quad \mbox{with}\quad r\in \R.
\end{eqnarray}

That is, for different values of $r$ we obtain different
self-adjoint extensions. Here a very natural extension is obtained
choosing $r=0$, that is,  imposing $\Psi(0)=0$. Physically, this is
equivalent to assume that at $a=0$ there is a infinite potential
barrier (in the same way that for no-relativistic one-dimensional
barrier problems), then the existence of a solution all the time is
guarantied because when the scale factor decreases to zero, at some
finite time, the potential barrier forces it to grow. Moreover, this
assumption  explains why the Heisenberg picture fails to work,
because in the Heisenberg picture the boundary conditions do not
appear, and the effective scale factor has the freedom to take all
the values in $\R$, in particular, $0$ or negative values. We can
conclude that if we want to work in Heisenberg picture we must
introduce some kind of potential barriers that prevent that the
effective scalar factor takes negative values.

Once we have obtained a self-adjoint extension we apply the
effective formulation (\ref{1a}) to the problem
\begin{eqnarray}\label{a35}
i\hbar\partial_t\Phi(t)=\frac{\h^2}{2\gamma^2a}\partial^2_{a}\Phi(t)+\rho_0a_0^3\Phi(t),\end{eqnarray}
with the additional conditions $ \Phi(t^*)=\Psi, \,\
\langle\hat{H}_{SA}\rangle_{\Psi}=0, \,\ ||\Psi||=1,$ that gives us
an strongly continuous unitary one-parameter group defined on
${\mathcal L}^2((0,\infty),ada)$ (Stone's theorem), namely
$e^{-\frac{i}{\hbar}\hat{H}_{SA}t}$. The solution of our problem can
be written as $\Phi(t)=e^{-\frac{i}{\hbar}\hat{H}_{SA}(t-t^*)}\Psi$
for all $\Psi\in D_{\hat{H}_{SA}}$ satisfying
$\langle\hat{H}_{SA}\rangle_{\Psi}=0$ and  $||\Psi||=1$. As an
example of initial condition, if $r=0$, one can take
\begin{eqnarray}\label{a36}
\Psi(a)\equiv
\frac{a^{-1}}{(\sigma\pi)^{1/4}}e^{-\frac{\ln^2(a/\bar{a})}{2\sigma}}
e^{\frac{i}{\hbar}(\ln(a/\bar{a}) p^*)},\quad\mbox{with}\quad
p^*=-\sqrt{2\rho_0(a_0\bar{a})^3\gamma^2e^{-\frac{9}{4}\sigma}-\hbar^2(25/4+1/(2\sigma))}.
\end{eqnarray}
For this initial state, the effective scale factor
$a_{eff}(t)=\langle \hat{a}\rangle_{\Phi(t)}$ grows forever for
$t>t^*$ in the similar way to the classical one (the classical limit
holds far of the turning point $a=0$). For $t<t^*$ the effective
scale factor decreases to zero, but at some finite time it bounces,
due to the potential barrier, and then it grows to infinity.

Finally, we study the case $\omega=1/3$ (radiation). The Hamiltonian
is
$\hat{H}=\frac{\h^2}{2\gamma^2a}\partial^2_{a}+\frac{\rho_0a_0^4}{a}$,
and the solutions of the equation $\hat{H}\Phi=\pm i\rho_0a_0^3\Phi$
are the Airy's functions $\Phi_{1,\pm}\equiv Ai(\beta_{\pm}( a\mp
ia_0))$ and $\Phi_{2,\pm}\equiv Bi(\beta_{\pm}( a\mp ia_0))$, where
$\beta_{\pm}\equiv
\left(\frac{2\gamma^2\rho_0a_0^3}{\hbar^2}\right)^{1/3}e^{\pm
i\pi/6}$.

In this case the dimension of both deficiency subspaces is $1$, then
as the dust matter case, $\hat{H}$ has infinitely many self-adjoint
extensions parametrized by an unitary $1\times 1$ matrix, and the
self-adjoint extensions are determined,  once again, by the boundary
condition $\Psi(0)= r\Psi'(0)$, with $r\in \R$. Now an initial
condition for our effective formulation, that exhibits the same
behavior as above for the effective scale factor, is given by the
function
\begin{eqnarray}\label{a37}
\Psi(a)\equiv
\frac{a^{-1}}{(\sigma\pi)^{1/4}}e^{-\frac{\ln^2(a/\bar{a})}{2\sigma}}
e^{\frac{i}{\hbar}(\ln(a/\bar{a}) p^*)}, \quad\mbox{with}\quad
p^*=-\sqrt{2\rho_0(a^2_0\bar{a})^2\gamma^2e^{-2\sigma}-\hbar^2(25/4+1/(2\sigma))}.
\end{eqnarray}

We finish this Appendix with the following remark. 
When the three-dimensional curvature is positive ($k=1$), the
Hamiltonian of the system is $H=-\frac{1}{2\gamma^2
a}p_a^2+\rho(a)a^3-\frac{1}{2}\gamma^2c^2a$. Then the Hamiltonian
constraint restricts the value of the scalar factor into the
interval $(0,A)$ with $A=
\left({2\rho_0a_0^{3(\omega+1)}}/(c\gamma)^2\right)^{\frac{1}{3\omega+1}}$,
and this say us that we must take as  Hilbert space, the space
${\mathcal L}^2\left((0,A),ada\right)$. Now for $\omega\leq 1$,
$a=0$ is a regular singular point of the ordinary differential
equation $\hat{H}\Phi=z_{\pm}\Phi$, then applying the Frobenius
method we can deduce that there exist two independent solutions of
the differential equation, consequently both deficiency indices are
$2$, because the domain $(0,A)$ is finite (excepts for
$\omega=-1/3$). Then
 the self-adjoint extensions are parametrized by an unitary
$2\times 2$ matrix, and the more natural boundary condition is to
assume that the wave-functions vanish at two boundary points.
Physically this means that the scale factor is confined in a very
deep well potential, and we have an oscillating universe whose
effective scalar factor never vanishes.

\section{Appendix C:  QFT in curved space-time from the effective formulation}

For the flat FRW universe, the action that describes  a massive
scalar field  conformally coupled with gravity in the presence of a
barotropic fluid , is given by
\begin{eqnarray}
S=\int_{\R}dt\int_{[0,L]^3}d\vec{x}\left[-\frac{\gamma^2}{2}\dot{a}^2a-\rho_0\left(a/a_0\right)^{-3(\omega+1)}a^3
+a^3{\mathcal L}_{\phi}\right],
\end{eqnarray}
with ${\mathcal L}_{\phi}=\frac{1}{2\hbar
c^3}\dot{\phi}^2-\frac{1}{2\hbar c
a^2}(\nabla\phi)^2-\frac{m^2c}{2\hbar^3}\phi^2-\frac{1}{12\hbar
c^3}R^2\phi^2$ where $R=\frac{6}{a^2}(\dot{a}^2+a\ddot{a})$ is the
scalar curvature. (Note that in this Appendix $\phi$ has energy
units).
Integrating with respect $\vec{x}$ and expanding $\phi$ in Fourier
series ($\phi=\sum_{\vec{k}\in\Z^3}\phi_{\vec k}e^{2\pi i\frac{\vec
k.\vec x}{L}}$) one obtains, $S=\int_{\R} L(t) dt$, with
\begin{eqnarray}
L(t)=L^3\left\{-\frac{\gamma^2}{2}\dot{{a}}^2{a}-\rho_0\left({a}/{a}_0\right)^{-3(\omega+1)}{a}^3
+{a}^3\sum_{{\vec k}\in\Z^3}{\mathcal L}_{\phi_{\vec k}}\right\},
\end{eqnarray}
where ${\mathcal L}_{\phi_{\vec k}}=\frac{1}{2\hbar
c^3}\dot{\phi}_{\vec k}^2-\frac{1}{2\hbar c a^2}\frac{4\pi^2|{\vec
k}|^2}{L^2}\phi_{\vec k}^2-\frac{m^2c}{2\hbar^3}\phi_{\vec
k}^2-\frac{1}{12\hbar c^3}R^2\phi_{\vec k}^2$

Using now the conformal time  $d\eta\equiv \frac{ct_{p}}{{a}}dt$, ($t_p$
being the Planck time) and defining the function $\psi_{\vec
k}=\sqrt{\frac{4\pi t_{p}}{3\hbar}}\frac{a}{c}\phi_{\vec k}$ we
obtain $L(t)dt\equiv \frac{3L^3}{4\pi}\tilde{L}(\eta)d\eta$, with
\begin{eqnarray}
\tilde{L}(\eta)=-\frac{\hbar}{2t_{p}}\left(\frac{{a}'}{c}\right)^2-\tilde{\rho}_0\left({{a}}/{a}_0\right)^{-3(\omega+1)}\frac{{a}^4}{l_{p}}
+\frac{1}{2}\sum_{{\vec k}\in\Z^3}\left((\psi')_{\vec
k}^2-\frac{1}{t_{p}^2 }\left[\frac{4\pi^2|{\vec
k}|^2}{L^2}+\left(\frac{{a}}{l_c}\right)^2\right] \psi_{\vec
k}^2\right),
\end{eqnarray}
where we have introduced the Compton wavelenght $l_c\equiv
\frac{\hbar}{mc}$, and we have defined
$\tilde{\rho}_0=\frac{4\pi}{3}{\rho}_0$.
The important remark should be made that 
in this Lagrangian we have suppressed the terms
$-\frac{3}{8\pi}\left(\frac{{a}'}{{a}}\psi_{\vec k}^2\right)'$.

The conjugate momenta are $ p_{{a}}=
-\frac{\hbar}{l_{p}}\frac{{a}'}{c}$, $p_{\psi_{\vec k}}= \psi_{\vec
k}'$, and the Hamiltonian is given by
\begin{eqnarray}
\tilde{H}(\eta)=-\frac{1}{2m_{p}}p_{{a}}^2+U({a})
+\frac{1}{2}\sum_{{\vec k}\in\Z^3}\left(p_{\psi_{\vec
k}}^2+\omega_{\vec k}^2({a}) \psi_{\vec k}^2\right),
\end{eqnarray}
where $m_p$ is the Planck mass and
\begin{eqnarray*}
U({a})\equiv\tilde{\rho}_0\left({{a}}/{a}_0\right)^{-3(\omega+1)}\frac{{a}^4}{l_{p}},
\qquad \omega_{\vec k}^2({a})\equiv\frac{1}{t_{p}^2
}\left[\frac{4\pi^2|{\vec
k}|^2}{L^2}+\left(\frac{{a}}{l_c}\right)^2\right].
\end{eqnarray*}

The quantum theory is obtained making the replacement
$p_{{a}}\longrightarrow -i\hbar\partial_{{a}}$ and $p_{\psi_{\vec
k}}\longrightarrow -i\hbar\partial_{\psi_{\vec k}}$. Then the
quantum hamiltonian is given by
\begin{eqnarray}
\hat{\tilde{H}}=\frac{\hbar^2}{2m_{p}}\partial^2_{{a}^2}+ U({a})
+\hat{{ H}}_m({a},\psi),
\end{eqnarray}
where  the matter hamiltonian is $\hat{{ H}}_m({a},\psi)=\sum_{{\vec
k}\in\Z^3} \left(\hbar\omega_{\vec k} \hat{A}_{\vec
k}^{\dagger}\hat{A}_{\vec k}+\frac{1}{2}\hbar\omega_{\vec
k}\right)$, where we have introduced the creation and anihilation
operators
\begin{eqnarray}
\hat{A}_{\vec k}^{\dagger}\equiv \frac{1}{\sqrt{2\hbar\omega_{\vec
k}}}(-\hbar\partial_{\psi_{\vec k}}+\omega_{\vec k}\psi_{\vec
k});\qquad \hat{A}_{\vec k}\equiv \frac{1}{\sqrt{2\hbar\omega_{\vec
k}}}(\hbar\partial_{\psi_{\vec k}}+\omega_{\vec k}\psi_{\vec k}).
\end{eqnarray}

Now, we show how one can obtain the QFT in curved space-time from the
WDW equation. If we consider the matter field as a small
perturbation, we look for solutions of the WDW equation with the
form $\Phi({a},\psi)=\Psi({a})\chi({a},\psi)$. After substitution in
the WDW equation we obtain:
\begin{eqnarray}
\left[\frac{\hbar^2}{2m_{p}}\partial^2_{{a}^2}\Psi+U({a})\Psi\right]\chi
+\left[\Psi\frac{\hbar^2}{2m_{p}}\partial^2_{{a}^2}\chi+\frac{\hbar^2}{m_{p}}\partial_{{a}}\Psi\partial_{{a}}\chi+
\Psi\hat{{ H}}_m\chi \right]=0.
\end{eqnarray}
We assume at this point that $\Psi$ is the solution of the equation
\begin{eqnarray}
-\frac{\hbar^2}{2m_{p}}\partial^2_{{a}^2}\Psi-U({a})\Psi=0,
\end{eqnarray}
and we make the change $\Psi=e^{-\frac{i}{\hbar}S}$, then we obtain
the system
\begin{eqnarray}\left\{\begin{array}{ccc}
\frac{(\partial_{{a}} S)^2}{2m_{p}}-U({a})+\frac{i\hbar}{2m_{p}}\partial_{{a}^2}^2S&=&0\\
\frac{\hbar^2}{2m_{p}}
\partial^2_{{a}^2}\chi
-i\hbar\frac{\partial_{{a}}
S}{m_{p}}\partial_{{a}}\chi+\hat{{H}}_m\chi&=&0.\end{array}\right.
\end{eqnarray}

To solve this equations we neglect, as Rubakov does \cite{r84}, the
second derivative with respect to ${a}$, then we obtain the system
\begin{eqnarray}\left\{\begin{array}{ccc}
\frac{(\partial_{{a}} S)^2}{2m_{p}}-U({a})&=&0\\
-i\hbar\frac{\partial_{{a}} S}{m_{p}}\partial_{{a}}\chi+\hat{{
H}}_m\chi&=&0.\end{array}\right.
\end{eqnarray}
The first equation is the classical Hamilton-Jacobi equation, and
the second one is the quantum Schr\"odinger equation that can be
solved choosing as solution of the Hamilton-Jacobi equation
$S({a})=\int^{{a}}_{0}\sqrt{2m_{p}U({a})}d{a}$, and introducing the
conformal time $\frac{d{a}}{d\tau}\equiv\frac{\partial_{
a}S}{m_{p}}$, then
 the Schr\"odinger equation becomes
 $i\hbar\partial_{\tau}\chi=\hat{H}_m(a(\tau),\psi)\chi$.

Finally, we device a method to obtain the QFT in curved space-time from
the effective equation
$i\hbar\partial_{\eta}\Phi=\hat{\tilde{H}}\Phi$. Assuming that the
matter field is an small perturbation, we look for solutions of the
form $\Phi({a},\psi;\eta)=\Psi({a};\eta)\chi(\psi;\eta)$ where
$\Psi$ is the solution of the equation
\begin{eqnarray}
i\hbar\partial_{\eta}\Psi=\frac{\hbar^2}{2m_{p}}\partial^2_{{a}^2}\Psi+U({a})\Psi,
\end{eqnarray}
and we assume that $\Psi$ is a function concentrated around a
classical solution, namely ${a}_c(\eta)$, of the following equation
\begin{eqnarray}\label{a50}
-\frac{1}{2m_{p}}p_{{a}}^2+U({a})=0.
\end{eqnarray}
By inserting $\Phi$ in the effective equation one obtains $\Psi
i\hbar\partial_{\eta}\chi=\Psi\hat{H}_m({a},\psi)\chi$, and since
$\Psi$ is concentrated around the classical solution, one can
approximate $\Psi\hat{H}_m({a},\psi)$ by
$\Psi\hat{H}_m({a}_c(\eta),\psi)$, and then one obtains
$i\hbar\partial_{\eta}\chi=\hat{H}_m({a}_c(\eta),\psi)\chi$.

We end with a last remark.
From the effective formulation it's not difficult to obtain the
semi-classical Einstein equations. Effectively, starting with the
condition $\langle\hat{\tilde{H}}\rangle_{\Phi}=0$, if we take the
wave function used above (now picked around $a_c+\delta a_c$), one
approximately obtain
\begin{eqnarray}
-\frac{1}{2m_{p}}p_{{a_c+\delta a_c}}^2+U({a_c+\delta
a_c})+\langle\hat{H}_m({a}_c(\eta)+\delta
a_c(\eta),\psi)\rangle_{\chi, ren}=0,
\end{eqnarray}
where the quantity $\langle\hat{H}_m({a}_c(\eta)+\delta
a_c(\eta),\psi)\rangle_{\chi}$ has been renormalized.

Since $a_c$ is solution of the equation (\ref{a50}), one also
obtains, in the linear approximation, the following back-reaction
equation:
\begin{eqnarray}
-\frac{\hbar}{cl_{p}}a'_c(\delta a_c)'+U'(a_c)\delta
a_c+\langle\hat{H}_m({a}_c(\eta),\psi)\rangle_{\chi, ren}=0.
\end{eqnarray}
Finally, observe that the derivation of the semi-classical Einstein
equation from the WDW one is not a completely clear case (see for
example \cite{ha87}).

 \vspace{1cm}

{\bf Acknowledgements:}

 This investigation has been supported in part by MEC
(Spain), projects MTM2005-07660-C02-01 and FIS2006-02842, and by
AGAUR (Generalitat de  Ca\-ta\-lu\-nya), contract 2005SGR-00790
and grant DGR2008BE1-00180. Part of EE's research was performed
while on leave at Department of Physics and Astronomy, Dartmouth
College, 6127 Wilder Laboratory, Hanover, NH 03755, USA.

\end{document}